\documentclass{article}%
\usepackage[utf8]{inputenc}
\usepackage[T1]{fontenc}
\usepackage{lmodern}
\usepackage{amsmath}
\usepackage{amsfonts}
\usepackage{amssymb}
\usepackage{graphicx}
\usepackage{cite}
\usepackage{hyperref}
\usepackage{authblk} 
\usepackage{orcidlink}
\title{Time-Dependent Dunkl-Schrödinger Equation \\
with an Angular-Dependent Potential}

\author[1]{B. C. L\"{u}tf\"{u}o\u{g}lu \orcidlink{0000-0001-6467-5005}\thanks{bekir.lutfuoglu@uhk.cz (Corresponding author)}}
\author[2,3]{A. Benchikha \orcidlink{0009-0003-9848-0822}\thanks{benchikha4@yahoo.fr/amar.benchickha@umc.edu.dz}}
\author[2]{B. Hamil \orcidlink{0000-0002-7043-6104} \thanks{hamilbilel@gmail.com/bilel.hamil@umc.edu.dz}}
\author[4,5]{B. Khantoul \orcidlink{0000-0001-9012-4864} \thanks{boubakeur.khantoul@univ-constantine3.dz}}

\affil[1]{ Department of Physics, Faculty of Science, University of Hradec Kralove, \quad\quad\qquad
Rokitanskeho 62/26, Hradec Kralove, 500 03, Czech Republic.}
\affil[2]{Laboratoire de Physique Mathématique et Subatomique, Faculté des Sciences Exactes, Université Constantine 1 Frères Mentouri, Constantine, Algeria.}
\affil[3]{Département de EC, Faculté de SNV, Université Constantine 1, Constantine, Algeria.}
\affil[4]{Department of Process Engineering, University of Constantine 3 - Salah Boubnider, Algeria.}
\affil[5]{Theoretical Physics Laboratory, Department of Physics, University of Jijel, Jijel, Algeria.}

\date{}
\begin{document}

\maketitle

\begin{abstract}
In this manuscript, we investigate the analytical solution of the time-dependent Schrödinger equation for a harmonic oscillator with time-dependent mass and frequency, coupled with angular-dependent potential energy by utilizing the Dunkl derivatives. To obtain the solution, we employ the Lewis-Riesenfeld invariant methodology. Our approach broadens the scope of quantum mechanical analyses, offering exact solutions and new insights into dynamic quantum systems under varying conditions.
\end{abstract}

\section{Introduction}

The Schrödinger equation represents a fundamental tenet of quantum mechanics, offering crucial insights into the behavior of quantum systems.  The Schrödinger equation has been the subject of extensive investigation with regard to a variety of potentials, resulting in the discovery of exact solutions and the formulation of valuable approximations in both stationary and time-dependent forms. One such area of interest is the harmonic oscillator potential, which serves as a fundamental model for understanding the vibrational modes in molecules, quantum field theories, and more \cite{1,2,3,4,5,6,7,8}. In recent years, significant progress has been made in various disciplines, including nanotechnology, charged particle beams in accelerators, and quantum optics, using solutions of the time-dependent Schrödinger equations. These progresses have enabled the modeling and analysis of systems with dynamically varying parameters, facilitating the exploration of dynamic quantum phenomena, including state evolution, quantum transport, and coherent control \cite{Nitzan}. However, profound mathematical difficulties limit further developments.



In order to overcome the existing difficulties, new approaches and methods have been proposed as alternatives to the existing traditional methods. Among these, Dunkl-quantum mechanics has attracted increasing interest, especially in the last decade, as it offers parity-dependent solutions \cite{Genest20131, Genest20132, Genest20133, Genest20141, Genest20142, Vincent3,  Isaac2016, Salazar2017, Salazar2018, Mota20181, Sargol2018, Hassanabadi2019, Ghazouani2020, Ojeda2020,  Mota20211, Mota20212, Bilel20221, Merad2021, Merad2022, Dong2021, Bilel20222, Halberg2022, Mota20221, Samira2022, Najafizade1, Najafizade2, Dong2023, Rouabhia2023, Samira2023, Ballesteros2024, Bouguerne2024, Bilel20223, Hassan2021, Fateh20231, Fateh20232, Hocine20241, Hocine20242, Ghazouani2023, ChungEPL2023, Sedaghatnia2023, Junker2023, Quesne2023, Quesne2024, Mota20241, Mota20242, Schulze20241, Schulze20242, Schulze20243, Benzair2024, Junker2024, Benchikha20241, Benchikha20242 }. The fundamental tenets of Dunkl quantum mechanics are founded upon the utilization of the Dunkl operators, which are derivative-difference operators associated with finite reflection groups \cite{Dunkl1989}, in lieu of the conventional partial derivative operator.  The rationale behind this utilization can be traced back to the mid-twentieth century, where initially Wigner \cite{Wigner1950} and subsequently Yang \cite{Yang1951} sought to derive commutation relations from the equations of motion. They demonstrated that, under specific constraints, this was feasible within a well-defined Hilbert space and with a deformed Heisenberg algebra. Not long after that Green employed the deformed algebra to propose a generalized field quantization, wherein he introduced novel concepts, including para-bosons, fermions, and statistics \cite{Green1953}. This approach proved instrumental in introducing color degrees of freedom and quantum chromodynamics \cite{Greenberg1964}. 

In the literature, the Dunkl operator with the Wigner deformation parameter, $\mu_{j}$, and the reflection operator, $\hat{R}_{j}$, is represented as follows \cite{Dunkl2014}:
\begin{equation}
\hat{D}_{j}=\frac{\partial}{\partial x_{j}}+\frac{\mu_{j}}{x_{j}}\left(
1-\hat{R}_{j}\right)  ,  \quad \quad \text{$j=1,2,3.$}\label{1}%
\end{equation}
and differs from the Yang operator by its second term. Here, the reflection operator obeys the following rules
\begin{equation}
\hat{R}_{j}f(x_{j})=f(-x_{j}),\quad\hat{R}_{i}\hat{R}_{j}=\hat{R}_{j}\hat
{R}_{i},\quad\hat{R}_{i}x_{j}=-\delta_{ij}x_{j}\hat{R}_{i},\quad\hat{R}%
_{i}\frac{\partial}{\partial x_{j}}=-\delta_{ij}\frac{\partial}{\partial
x_{j}}\hat{R}_{i},\label{2}%
\end{equation}
and its impact within the Dunkl operator produces different outcomes depending on whether the function it operates on is even or odd. Essentially, this perspective has led to the increased use of the Dunkl operator as a replacement for the conventional partial derivative operator {\cite{Chakra1994, Mikn1, Mikn2, Lapointe1996, Hikami1996, Kakei1996, Mik1, Mik2, Gamboa1999, Klishevich2001, Mikn3, Rodrigues2009, Horvathy2010, Ubriaco2014,  Jang2016, Mikn4, Chung20202, Luo2020, Chung2021, Ghazouani2021, AHalberg2022, MotOjeda2022, Dong2022, Sedaghatnia20232, nef1, nef7, nef8}.}


In physics, angle-independent potential functions are widely used due to their symmetric properties and the ease of obtaining analytical solutions. However, some other studies show that angular potentials bring additional richness to the system dynamics beyond central potentials, albeit more complexity through angular momentum operators {\cite{Hartman1972, Hautot1973, Kibler1984, Kibler1987, Quesne1988, Draganascu1992, Chang1999, Chen2002, Alhaidari2005, Berkdemir2008, Berkdemir20091, Berkdemir20092, Zhang2010, Agboola2011, Afs2015, Li2017, You2018, Sobhani2019, Moumni2020, Bouchefra2022, Ahmadov2023, Arabsaghari2024, nef2, nef3, nef4, nef5, nef6}}. In the Dunkl formalism, the Dunkl-angular operators diverge from the conventional ones, particularly in polar and spherical coordinates \cite{Vincent3, Ghazouani2020}. This fact underscores the significance of the angular dependence of the potential, particularly in two- and three-dimensional Dunkl-quantum mechanical problems. A thorough examination of the problems associated with angular-dependent potentials in the presence of Dunkl operators is lacking in the existing literature. Only very recently, in a stationary non-relativistic system Arabsaghari et al obtained a closed-form solution for an angular-dependent potential and determined the total energy change using the Hellman-Feynman theorem, which aids in calculating expectation values and analyzing energy shifts  \cite{Arabsaghari2024}. However, the dynamics of a time-dependent quantum mechanical system with an angular-dependent potential remain unexplored in the context of the Dunkl formalism. 

This motivates us to derive the exact solution of the three-dimensional time-dependent Schrödinger equation (TDSE) with harmonic oscillators, time-dependent mass and frequency, and angular-dependent potential energy in the Dunkl formalism and to explore the additional effects on some physical quantities deduced by the Dunkl operators. It is worth emphasizing that the extension of the Schr\"{o}dinger equation for a harmonic oscillator with time-dependent mass and frequency could be particularly relevant to realistic physical scenarios where system parameters vary with time, such as in driven quantum systems or time-dependent fields. In addition, the presence of a time-dependent angular potential would add further complexity but richness to the dynamics of the system. To overcome the mathematical obstacles, we consider the use of an invariant method, in particular the Lewis-Riesenfeld invariant method, which has been shown to be a powerful tool for solving the TDSE \cite{Lewis1969}. The aforementioned method has recently been successfully employed in the context of time-dependent harmonic oscillator problems, specifically in the context of time-dependent Dunkl-Schr\"{o}dinger equation
\cite{Benchikha20242}. 

Consequently, in the current manuscript using the Lewis-Riesenfeld invariant method, we solve the three-dimensional TDSE  for a harmonic oscillator with time-dependent mass and frequency, coupled with a time-dependent angular potential in the presence of the Dunkl formalism. This paper is structured as follows: Section \ref{sec2} presents the construction of the Dunkl-Hamiltonian for the TDSE with a specified time-angular potential. Section \ref{sec3} addresses the solution of the equation through the application of the invariant method and separation of variables, complemented by other analytical techniques. Section \ref{sec4} offers concluding remarks, emphasizing the significance of the findings and potential avenues for future research.

\section{Time-Dependent Dunkl-Schr\"{o}dinger Equation  with Angular-Dependent Potential}\label{sec2}

In this section, we delve into the exact solutions of the time-dependent
Dunkl-Schr\"{o}dinger equation with an angular-dependent potential:%
\begin{equation}
\mathcal{H}\left(  t\right)  \Psi\left(  \overrightarrow{r},t\right)
=i\frac{\partial}{\partial t}\Psi\left(  \overrightarrow{r},t\right)
.\label{4}%
\end{equation}
The Hamiltonian for this system is given by:
\begin{equation}
\mathcal{H}=-\frac{1}{2M(t)}\left(  \hat{D}_{1}^{2}+\hat{D}_{2}^{2}+\hat
{D}_{3}^{2}\right)  +\frac{1}{2}M(t)\omega^{2}(t)(x_{1}^{2}+x_{2}^{2}%
+x_{3}^{2})+\frac{U(\theta,\varphi)}{M(t)(x_{1}^{2}+x_{2}^{2}+x_{3}^{2}%
)},\label{5}%
\end{equation}
Here, $M(t)$ and $\omega(t)$ denote the time-dependent mass and frequency,
respectively, while $U(\theta,\varphi)$ represents the angular-dependent
potential \cite{Ang3,Ang4,Ang5,Ang6}
\begin{equation}
U(\theta,\varphi)=\frac{b}{2}\cot^{2}\theta+\frac{a}{\sin^{2}\theta\sin
^{2}\varphi},\label{6}%
\end{equation}
with real coupling constants $a$ and $b$. 

To facilitate the solution process, we use the spherical coordinates 
\begin{equation}
x_{1}=r\cos\varphi\sin\theta,\quad x_{2}=r\sin\varphi\sin\theta,\quad
x_{3}=r\cos\theta, \label{7}%
\end{equation}
and  transform the Hamiltonian into the following form 
\begin{equation}
\mathcal{H}=\frac{1}{2M(t)}\mathbf{p}^{2}+\frac{1}{2}M(t)\omega^{2}%
(t)r^{2}+\frac{\hbar^{2}\delta(\delta-1)+2U(\theta,\varphi)}{2M(t)r^{2}%
},\label{8}%
\end{equation}
with the momentum operator in three dimensions \cite{Benchikha20242}
\begin{equation}
\mathbf{p}^{2}=\mathcal{P}_{r,\delta}^{2}+\frac{L_{D}^{2}}{r^{2}},\label{9}%
\end{equation}
and the operator,  $\mathcal{P}_{r,\delta}$, \begin{equation}
\mathcal{P}_{r,\delta}=\frac{\hbar}{i}\left(  \frac{\partial}{\partial
r}+\frac{\delta}{r}\right)  .\label{10}%
\end{equation}
Here, the Dunkl-angular momentum operator $L_{D}^{2}$ reads:
\begin{align}
L_{D}^{2} &  =-\hbar^{2}\left\{  \frac{\partial^{2}}{\partial\theta^{2}%
}+2\left[  \left(  \frac{1}{2}+\mu_{1}+\mu_{2}\right)  \cot\theta-\mu_{3}%
\tan\theta\right]  \frac{\partial}{\partial\theta}-\frac{\mu_{3}}{\cos
^{2}\theta}(1-\hat{R}_{3})\right.  \nonumber\\
&  +\frac{1}{\sin^{2}\theta}\left.  \left(  \frac{\partial^{2}}{\partial
\varphi^{2}}+2\left(  \mu_{2}\cot\varphi-\mu_{1}\tan\varphi\right)
\frac{\partial}{\partial\varphi}-\frac{\mu_{1}}{\cos^{2}\varphi}%
(1-\hat{R}_{1})-\frac{\mu_{2}}{\sin^{2}\varphi}(1-\hat{R}_{2})\right)  \right\}
,\label{11}%
\end{align}
where
\begin{equation}
\delta=\mu_{1}+\mu_{2}+\mu_{3}+1, \quad \text{and} \quad [P_{r,\delta},r]=i\hbar.
\end{equation}
Now, we express the time-dependent Schrödinger equation which we aim to solve
\begin{equation}
i\hbar\frac{\partial}{\partial t}\psi(r,\theta,\varphi,t)=\left[  \frac
{1}{2M(t)}\mathbf{p}^{2}+\frac{1}{2}M(t)\omega^{2}(t)r^{2}+\frac{\hbar
^{2}\delta(\delta-1)}{2M(t)r^{2}}+\frac{U(\theta,\varphi)}{M(t)r^{2}}\right]
\psi(r,\theta,\varphi,t), \label{12}%
\end{equation}
with the angular-dependent potential given in Eq. \eqref{6}.

\subsection{Exact solution}\label{sec3}

To solve  Eq. \eqref{12}, we employ the Lewis-Riesenfeld invariant approach. This method allows us to derive exact solutions for the time-dependent Schrödinger equation, broadening our understanding of quantum systems under dynamic conditions. According to this method, the Hamiltonian and the  invariant, $I$,  should satisfy the Lewis-Riesenfeld invariant equation
\begin{equation}
\frac{dI(t)}{dt}=\frac{\partial I(t)}{\partial t}+\frac{1}{i\hbar
}\left[I(t),\mathcal{H}(t)\right]=0.\label{13}%
\end{equation}
This allows us to relate the solution of the time-dependent Schrödinger equation, $\Psi(\vec{r},t)$, to the solution of the invariant's eigenvalue problem
\begin{equation}
I(t)\Upsilon(\vec{r},t)=E_{n,l,m}\Upsilon(\vec{r},t),\label{14}%
\end{equation}
with a phase factor
\begin{equation}
\Psi(\vec{r},t)=e^{i\eta(t)}\Upsilon(\vec{r},t),\label{15}%
\end{equation}
where $\eta(t)$ can be determined from the equation
\begin{equation}
\hbar\frac{d}{dt}\eta_{n}(t)=\langle\Upsilon_{n}(\vec{r},t)|i\hbar
\frac{\partial}{\partial t}-H|\Upsilon_{n}(\vec{r},t)\rangle.\label{16}%
\end{equation}
To construct the exact invariant for the quantum system, we then introduce the generators $\left\{  T_{1},T_{2},T_{3}\right\} $ in the following explicit forms:
\begin{align}
T_{1} &  =\mathbf{p}^{2}+\left(  \frac{\hbar^{2}\delta(\delta-1)+2U(\theta
,\varphi)}{r^{2}}\right)  ,\\
T_{2} &  =r^{2},\\
T_{3} &  =\frac{1}{2}\left(  r\mathcal{P}_{r,\delta}^{2}+\mathcal{P}%
_{r,\delta}^{2}r\right) ,
\end{align}
which verifies the following commutation relations:
\begin{equation}
\lbrack T_{1},T_{2}]=-2i\hbar T_{3},\quad\lbrack T_{2},T_{3}]=4i\hbar
T_{2},\quad\lbrack T_{1},T_{3}]=-4i\hbar T_{1}.\label{17}%
\end{equation}
Let us now proceed with the assumption that the invariant follows the following form:
\begin{equation}
I=\frac{1}{2}(\alpha T_{1}+\beta T_{2}+\gamma T_{3}). \label{18}%
\end{equation}
Subsequently,  utilizing Eq. \eqref{13},  we obtain a system of equations whose solution is contingent upon the real-time variation of the functions $\alpha$, $\beta$ and $\gamma$
\begin{eqnarray}
\alpha &=& \rho^{2},\\
\beta &=& \frac{1}{\rho^{2}}+M^{2}\dot{\rho}^{2},\\
\gamma &=& -M\rho\dot{\rho}.
\end{eqnarray}
Thus, the invariant takes the following explicit form
\begin{equation}
I=\frac{1}{2}\left[ \left( \frac{1}{\rho ^{2}}+M^{2}\dot{\rho}^{2}\right)
r^{2}+\rho ^{2}\left( \mathbf{p}^{2}+\frac{\hbar ^{2}\delta \left( \delta
-1\right) +2U\left( \theta ,\varphi \right) }{r^{2}}\right) -\rho \dot{\rho}%
M\left( r\mathcal{P}_{r,\delta }+\mathcal{P}_{r,\delta }r\right) \right] ,\label{19}%
\end{equation}
where $\rho$ satisfies the Ermakov-Pinney equation given by:
\begin{equation}
\ddot{\rho}+\frac{\dot{M}}{M}\dot{\rho}+\omega^{2}(t)\rho=\frac{1}{M^{2}%
\rho^{3}}.\label{20}%
\end{equation}
To solve the eigenvalue equation given in  Eq. \eqref{14}, we introduce the following unitary transformation
\begin{equation}
\Upsilon\left(  \overrightarrow{r}\right)  =S\mathcal{F}\left(\overrightarrow{r}\right)  =e^{\frac{iM\dot{\rho}}{2\hbar\rho}r^{2}}\mathcal{F}\left(  \overrightarrow{r}\right)  ,\label{21}%
\end{equation}
so the invariant transforms as
\begin{equation}
I\acute{}=\frac{1}{2}\left[ \rho ^{2}\left( \mathbf{p}^{2}+\frac{\hbar
^{2}\delta \left( \delta -1\right) +2U\left( \theta ,\varphi \right) }{r^{2}}%
\right) +\frac{1}{\rho ^{2}}r^{2}\right]. \label{22}%
\end{equation}
We then express the  multi-variable function $\mathcal{F}\left(  \overrightarrow{r}\right)$ as the multiplication of radial, polar and azimuthal functions
\begin{eqnarray}
\mathcal{F}(r,\theta,\varphi)= r^{-\delta }R\left(  r\right)\Theta\left(  \theta\right)  \Phi\left(  \varphi\right)  .
\end{eqnarray}
This assumption lets us decompose  the eigenvalue equation, Eq. \eqref{14}, into the following set of functions
\begin{eqnarray}
-\left[  \frac{\partial^{2}}{\partial\varphi^{2}}+2\left(  \mu_{2}\cot
\varphi-\mu_{1}\tan\varphi\right)  \frac{\partial}{\partial\varphi}-\frac
{\mu_{1}}{\cos^{2}\varphi}(1-\hat{R}_{1})-\frac{\mu_{2}}{\sin^{2}\varphi}%
(1-\hat{R}_{2})-\frac{2a}{\hbar^{2}\sin^{2}\varphi}\right]  \Phi\left(
\varphi\right)  &=& m^{2}\Phi\left(  \varphi\right) , \,\,\,\, \label{24} \\
-\left[ \frac{\partial^{2}}{\partial\theta^{2}}-2\left(\mu_{3}\tan\theta-\left(  \frac{1}%
{2}+\mu_{1}+\mu_{2}\right)  \cot\theta\right)
\frac{\partial}{\partial\theta}-\frac{m^{2}}{\sin^{2}\theta}-\frac{\mu_{3}}{\cos^{2}\theta}(1-\hat{R}_{3}%
)-b\cot^{2}\theta\right] \Theta\left(  \theta\right)  &=&\lambda\Theta\left(\theta\right) , \label{23} \\
-\rho ^{2}\left[ \frac{\partial ^{2}}{\partial r^{2}}-\frac{\delta \left(
\delta -1\right) +\lambda }{r^{2}}-\frac{1}{\hbar ^{2}}\frac{r^{2}}{\rho ^{4}%
}\right] R\left( r\right) =\frac{2E_{n,l,m}}{\hbar ^{2}}R\left( r\right) . \label{25}%
\end{eqnarray}
Here, $m^2$ and $\lambda = l(l+1)$ represent the separation constants, where $m$, $\lambda$, and $l$ are the standard magnetic quantum numbers.
The general form of the azimuthal eigenfunction reads:
\begin{equation}
\Phi\left( \varphi \right) =\mathcal{C}_{\varphi }\cos \left( \varphi
\right) ^{\frac{1-e_{1}}{2}}\sin \left( \varphi \right) ^{\frac{1}{2}-\mu
_{2}+\alpha }P_{n_{\varphi }}^{^{\left( \alpha ,\text{ }\mu _{1}-\frac{e_{1}%
}{2}\right) }\text{ }}\left( \cos 2\varphi \right) ,
\end{equation}
where
\begin{eqnarray}
\alpha =\sqrt{\left( \mu _{2}-\frac{e_{2}}{2}\right) ^{2}+\frac{2a}{\hbar^{2}}},
\end{eqnarray}
and $P_{n}^{\left(  p,q\right)}$ is the generalized Jacobi polynomial 
\cite{Ang7}. Here, $R_{1}\Phi_{m}\left(  \varphi\right)  =e_{1}\Phi_{m}\left(
\varphi\right)  $ and $R_{1}\Phi_{m}\left(  \varphi\right)  =e_{2}\Phi
_{m}\left(  \varphi\right)  ,$ where $e_{1}$ and $e_{1}$\ attain the values
$\pm1$. Besides, $n_{\varphi}$ is a positive integer related to the separation constant $m^{2}$ as follows
\begin{eqnarray}
m^{2}=\left[ 2n_{\varphi }+1+\mu _{1}-\frac{e_{1}}{2}+\alpha \right]
^{2}-\left( \mu _{1}+\mu _{2}\right) ^{2}.\label{27}%
\end{eqnarray}
For the action of the reflection operator, $R_{3}$, on the polar function
$\Theta\left(  \theta\right)  $, we employ $R_{3}\Theta\left(  \theta\right)
=e_{3}\Theta\left(  \theta\right)  $, where $e_{3}$ attains the value $\pm1$.
Thus, the polar equation $\Theta\left(  \theta\right)  $ takes the form:%

\begin{equation}
\Theta \left( \theta \right) =\mathcal{C}_{\theta }\cos \left( \theta
\right) ^{\frac{1-e_{3}}{2}}\sin \left( \theta \right) ^{\beta -\mu _{1}-\mu
_{2}}P_{n_{\theta }}^{\left( \beta ,\text{ }\mu _{3}-\frac{e_{3}}{2}\right) 
\text{ }}\left( \cos 2\theta \right) , \label{28c}
\end{equation}
where
\begin{eqnarray}
\beta =\sqrt{m^{2}+\frac{b}{\hbar ^{2}}+\left( \mu _{1}+\mu _{2}\right) ^{2}},
\end{eqnarray}
for a nonnegative integer $n_{\theta}$. Here, we see that the separation constant, $\lambda$, is interrelated in terms of $n_{\theta}$ with the following relation
\begin{eqnarray}
 \lambda =\left[ 2n_{\theta }+1+\mu _{3}-\frac{e_{3}}{2}+\beta \right] ^{2}-%
\frac{b}{\hbar ^{2}}-\left( \frac{1}{2}+\mu _{1}+\mu _{2}+\mu _{3}\right)
^{2}. \label{29}%
\end{eqnarray}
In order to eliminate the time-dependent function $\rho$ from the radial equation, we use the transformation $\frac{r}{\rho}=\varkappa$. We then rewrite Eq.  \eqref{25} as 
\begin{equation}
\left\{ \frac{\partial ^{2}}{\partial \mathcal{\varkappa }^{2}}+\frac{1}{%
\hbar ^{2}}\left[ 2E_{n,l,m}-\mathcal{\varkappa }^{2}-\frac{\hbar ^{2}\left(
\delta \left( \delta -1\right) +\lambda \right) }{\mathcal{\varkappa }^{2}}-%
\right] \right\} R\left( \mathcal{\varkappa }\right) =0.\label{30}%
\end{equation}
The solution to this equation is a straightforward matter of calculation, and the result is as follows
\begin{equation}
R\left( \mathcal{\varkappa }\right) =\mathcal{C}_{r}\mathcal{%
\varkappa }^{\sigma +\frac{1}{2}}e^{-\frac{\mathcal{\varkappa }^{2}}{2\hbar }%
}L_{n}^{\sigma }\left( \frac{\mathcal{\varkappa }^{2}}{\hbar }\right) .
\label{31}
\end{equation}
Here, $C_{r}$ is the normalization constant and the eigenvalue of the invariant reads: 
\begin{equation}
E_{n,l,m}=\hbar(2n+\sigma+1),\quad \text{where} \quad n=1,2,3...., \label{32}%
\end{equation}
and
\begin{equation}
\sigma=\sqrt{\frac{1}{4}+\delta\left(  \delta-1\right)  +\lambda}.\label{33}%
\end{equation}
To calculate the quantum phase $\eta\left(  t\right)  $, we substitute the Hamiltonian
\begin{equation}
\mathcal{H=}\frac{I}{M\rho^{2}}-\left(  \frac{1}{2M\rho^{4}}+\frac{M\dot{\rho
}^{2}}{2\rho^{2}}\right)  x^{2}-\frac{\dot{\rho}}{2\rho}\left(  r\mathcal{P}%
_{r,\delta}+\mathcal{P}_{r,\delta}r\right)  +\frac{M \omega^{2}x^{2}}{2}%
,\label{34}%
\end{equation}
 in Eq. \eqref{16}. We find
\begin{equation}
\hbar \dot{\eta}\left( t\right) =-\frac{E}{M\rho ^{2}}+\Big\langle 
\mathcal{F}\Big\vert \left. i\hbar \frac{\partial }{\partial t}-%
\frac{\dot{\rho}}{2\rho }\left( r\mathcal{P}_{r,\delta }+\mathcal{P}%
_{r,\delta }r\right) \right. \Big\vert \mathcal{F}\Big\rangle .\label{35}%
\end{equation}
By employing the unitary transformation, as defined in Eq. \eqref{21}, and taking into account the Ermakov-Pinney equation, as presented in Eq. \eqref{20}, one can perform straightforward calculations to obtain
\begin{equation}
\Big\langle \mathcal{F}\Big\vert \left. i\hbar \frac{\partial }{%
\partial t}-\frac{\dot{\rho}}{2\rho }\left( r\mathcal{P}_{r,\delta }+%
\mathcal{P}_{r,\delta }r\right) \right. \Big\vert \mathcal{F}%
\Big\rangle =0.\label{36}%
\end{equation}
Thus, 
\begin{equation}
\dot{\eta}\left( t\right) =-\frac{E}{\hbar M\rho ^{2}}, \label{37}%
\end{equation}
and hence, the phase factor becomes
\begin{equation}
\eta \left( t\right) =-\left( 2n+1+\sigma \right) \int_{0}^{t}\frac{dt^{,}}{%
M\left( t^{,}\right) \rho \left( t^{,}\right) ^{2}}. \label{38}%
\end{equation}
Finally, the total wave function reads:
\begin{eqnarray}
\Psi(r,\theta ,\varphi ,t) &=&\mathcal{C}_{r,\theta ,\varphi }\exp \left[ 
\frac{\left( iM\rho \dot{\rho}-1\right) r^{2}}{2\hbar \rho ^{2}}-i\left(
\left( 2n+1+\sigma \right) \int_{0}^{t}\frac{dt\prime }{M\left( t^{\prime
}\right) \rho \left( t^{\prime }\right) ^{2}}\right) \right] \frac{r^{\sigma
-\delta +\frac{1}{2}}}{\rho ^{\sigma -\delta +1}}L_{n}^{\sigma }\left( \frac{%
r^{2}}{\hbar \rho ^{2}}\right)   \notag \\
&\times& \cos \left( \varphi \right) ^{\frac{1-e_{1}}{2}}\sin \left( \varphi
\right) ^{\frac{1}{2}-\mu _{2}+\alpha }P_{n_{\varphi }}^{^{\left( \alpha ,%
\text{ }\mu _{1}-\frac{e_{1}}{2}\right) }\text{ }}\left( \cos 2\varphi
\right)   \notag \\
&\times& \cos \left( \theta \right) ^{\frac{1-e_{3}}{2}}\sin \left( \theta
\right) ^{\beta -\mu _{1}-\mu _{2}}P_{n_{\theta }}^{\left( \beta ,\text{ }%
\mu _{3}-\frac{e_{3}}{2}\right) \text{ }}\left( \cos 2\theta \right).
\end{eqnarray}
By making specific selections, the wave function can be normalized through 
\begin{equation}
\int_{0}^{2\pi}\int_{0}^{\pi}\int_{0}^{+\infty}r^{2+2\mu_{1}+2\mu_{2}+2\mu
_{3}}\left\vert \sin\left(  \theta\right)  \right\vert ^{2\mu_{1}+2\mu_{2}%
}\left\vert \cos\left(  \theta\right)  \right\vert ^{2\mu_{3}}\left\vert
\sin\left(  \varphi\right)  \right\vert ^{2\mu_{2}}\left\vert \cos\left(
\varphi\right)  \right\vert ^{2\mu_{1}}\psi_{n^{,}}\psi_{n}drd\theta
d\varphi=\delta_{n,n^{,}} .\label{41}%
\end{equation}

\section{Conclusion} \label{sec4}

It is a challenging task to identify quantum systems in nature that can be accurately described using the solutions of the time-independent Schrödinger equation. {Realistic} physical scenarios, where system parameters vary over time, such as in driven quantum systems or time-dependent fields, {are} more accurately examined with the solutions of the time-dependent Schrödinger equation.  The dynamics of these systems are strongly dependent on the potential energies under consideration. In this framework, the introduction of angular-dependent potential energies, in addition to radial ones, adds significant complexity and richness to the dynamics of the systems under study.

{The Dunkl derivative, with its differential-difference operators associated with finite reflection groups, enables solutions to be decomposed under parity, offering a powerful tool for analyzing quantum systems. Our study highlights the advantages of Dunkl operators in solving the time-dependent Schrödinger equation (TDSE), particularly for angular-dependent systems. These operators simplify the treatment of angular dynamics by incorporating reflection symmetries, enabling parity-dependent solutions and providing new insights into system behavior. All our findings are inherently tied to the parity and Wigner deformation factors, which play a central role in the Dunkl formalism.}

In light of the aforementioned evidence, in this study we explored the solutions of the time-dependent Dunkl-Schrödinger equation by integrating a harmonic oscillator with time-varying mass and frequency, coupled with an angular-dependent potential energy in three dimensions. More precisely, using the Lewis-Riesenfeld invariant method, we solved the TDSE for a harmonic oscillator with time-varying mass and frequency, coupled with an angular-dependent potential in the Dunkl formalism. {The physical implications of our findings reveal how angular-dependent potentials introduce anisotropies that influence observable quantities, such as energy spectra and wavefunction evolution. The interplay between Dunkl operator symmetry and angular dynamics is evident in the energy shifts and state transitions observed in our model, with potential applications in quantum optics, molecular physics, and systems driven by time-dependent fields.}

Our results emphasize the importance of considering angular-dependent potentials in the presence of Dunkl operators, a previously underexplored area in quantum mechanics. {Future research could extend this work by exploring other types of time-dependent potentials, analyzing higher-dimensional systems, and investigating applications in more complex quantum mechanical problems involving angular interactions and their observable effects.}

\section*{Acknowledgments}
{The authors are thankful to the anonymous reviewer for his/her constructive comments.} B. C. L. is grateful to Excellence project PřF UHK 2211/2023-2024 for the financial support.


\begin{thebibliography}{99}                                                                           
\bibitem {1} M. A. Markov, (Ed.),  \emph{Invariants and the Evolution of Nonstationary Quantum Systems}, (Nova Science Publishers, Commack, New-York, 1989).

\bibitem {2} C. I. Um, K. H. Yeon, T. F. George, \href{https://www.sciencedirect.com/journal/physics-reports/vol/362/issue/2}{Phys. Rep. \textbf{362}, 63 (2002) .}

\bibitem {3} M. Sebawe Abdalla,  P. G. L. Leach, \href{https://iopscience.iop.org/article/10.1088/0305-4470/36/49/005}{J. Phys. A \textbf{36}, 12205 (2003).} 

\bibitem {4} M. Sebawe Abdalla,  P. G. L. Leach, \href{https://iopscience.iop.org/article/10.1088/0305-4470/38/4/008}{J. Phys. A \textbf{38}, 881 (2005).} 

\bibitem {5} Y. Bouguerra, M. Maamache, A. Bounames, \href{https://link.springer.com/article/10.1007/s10773-006-9145-9}{Int. J. Theor. Phys. \textbf{45}, 1791 (2006).}

\bibitem {6} S. Menouar, M. Maamache, Y. Saadi, J. R. Choi, \href{https://iopscience.iop.org/article/10.1088/1751-8113/41/21/215303}{J. Phys. A: Math. Theor. \textbf{41}, 215303 (2008).} 

\bibitem {7} B. Khantoul, A. Bounames, M. Maamache, \href{https://link.springer.com/article/10.1140/epjp/i2017-11524-7}{Eur. Phys. J. Plus \textbf{132}, 258 (2017).}

\bibitem {8} B. Khantoul, A. Bounames, \href{https://ojs.cvut.cz/ojs/index.php/ap/article/view/8348}{Acta Polytech. \textbf{63}, 132 (2023).} 

\bibitem{Nitzan} A. Nitzan, \emph{Quantum dynamics using the time-dependent Schrödinger equation, Chemical Dynamics in Condensed Phases: Relaxation, Transfer, and Reactions in Condensed Molecular Systems}, (Oxford Academic Press, Oxford, 2024).  

\bibitem{Genest20131} V. X. Genest, M. E. H. Ismail, L. Vinet, A. Zhedanov, \href{https://doi.org/10.1088/1751-8113/46/14/145201}{J. Phys. A: Math. Theor. \textbf{46}, 145201 (2013).}

\bibitem{Genest20132} V. X. Genest, L. Vinet, A. Zhedanov, \href{https://doi.org/10.1088/1751-8113/46/32/325201}{J. Phys. A: Math. Theor. \textbf{46}, 325201 (2013).} 

\bibitem{Genest20133} V. X. Genest, J. M. Lemay, L. Vinet, A. Zhedanov, \href{https://doi.org/10.1088/1751-8113/46/50/505204}{J. Phys. A: Math. Theor. \textbf{46}, 505204 (2013).} 

\bibitem{Genest20141} V. X. Genest, M. E. H. Ismail, L. Vinet, A. Zhedanov, \href{https://doi.org/10.1007/s00220-014-1915-2}{Commun. Math. Phys. \textbf{329}, 999 (2014).}

\bibitem{Genest20142} V. X. Genest L. Vinet, A. Zhedanov, \href{https://doi.org/10.1088/1742-6596/512/1/012010}{J. Phys.: Conf. Ser. \textbf{512}, 012010 (2014).}

\bibitem{Vincent3} V. X. Genest, A. Lapointe, L. Vinet, \href{https://doi.org/10.1016/j.physleta.2015.01.023}{Phys. Lett. A, \textbf{379}, 923 (2015).}



\bibitem{Isaac2016} P. S. Isaac, I. Marquette, \href{https://doi.org/10.1088/1751-8113/49/11/115201}{J. Phys. A: Math. Theor. \textbf{49}, 115201 (2016).} 

\bibitem{Salazar2017} M. Salazar-Ram\'irez, D. Ojeda-Guill\'en, R. D. Mota, V. D. Granados, \href{https://doi.org/10.1140/epjp/i2017-11314-3}{Eur. Phys. J. Plus \textbf{132}, 39 (2017).}

\bibitem{Salazar2018} M. Salazar-Ram\'irez, D. Ojeda-Guill\'en, R. D. Mota, V. D. Granados, \href{https://doi.org/10.1142/S0217732318501122}{Mod. Phys. Lett. A \textbf{33}, 1850112 (2018).}


\bibitem{Sargol2018} S. Sargolzaeipor, H. Hassanabadi, W. S. Chung, \href{https://doi.org/10.1142/S0217732318501468}{Mod. Phys. Lett. A \textbf{33}, 1850146 (2018).}

\bibitem{Mota20181} R. D. Mota, D. Ojeda-Guill\'{e}n, M. Salazar-Ram\'{\i}rez, V. D. Granados, \href{https://doi.org/10.1016/j.aop.2019.167964}{Ann. Phys. \textbf{411}, 167964 (2019).}

\bibitem{Hassanabadi2019} W. S. Chung, H. Hassanabadi, \href{https://doi.org/10.1142/S0217732319501906}{Mod. Phys. Lett. A \textbf{34}, 1950190 (2019).}

\bibitem{Ghazouani2020} S. Ghazouani, S. Insaf, \href{https://doi.org/10.1088/1751-8121/ab4a2d}{J. Phys. A: Math. Theor. \textbf{53}, 035202 (2020).} 


\bibitem{Ojeda2020} D. Ojeda-Guill\'{e}n, R. D. Mota, M. Salazar-Ram\'{\i}rez, V. D. Granados, \href{https://doi.org/10.1142/S0217732320502557}{Mod. Phys. Lett. A \textbf{35}, 2050255 (2020).}

\bibitem{Hassan2021} H. Hassanabadi, M. de Montigny, W. S. Chung, \href{https://doi.org/10.1016/j.physa.2021.126154}{Physica A \textbf{580}, 126154 (2021).}

\bibitem{Mota20211} R. D. Mota, D. Ojeda-Guill\'{e}n, M. Salazar-Ramirez, V. D. Granados, \href{https://doi.org/10.1142/S0217732321500668}{Mod. Phys. Lett. A \textbf{36},  2150066 (2021).}

\bibitem{Dong2021} S. H. Dong, W. H. Huang, W. S. Chung, P. Sedaghatnia, H. Hassanabadi, \href{https://doi.org/10.1209/0295-5075/ac2453}{EPL \textbf{135}, 30006 (2021).}


\bibitem{Mota20212} R. D. Mota, D. Ojeda-Guill\'{e}n, M. Salazar-Ramirez, V. D. Granados, \href{https://doi.org/10.1142/S0217732321501716}{Mod. Phys. Lett. A \textbf{36},  2150171 (2021).}


\bibitem{Merad2021} A. Merad, M. Merad, \href{https://doi.org/10.1007/s00601-021-01683-4}{Few-Body Syst. \textbf{62}, 98 (2021).}


\bibitem{Merad2022} A. Merad, M. Merad, T. Boudjedaa, \href{https://doi.org/10.1142/S0217751X22500725}{Int. J. Mod. Phys. A \textbf{37}, 2250072 (2022).}

\bibitem{Bilel20221} B. Hamil, B. C. L\"{u}tf\"{u}o\u{g}lu, \href{https://doi.org/10.1007/s00601-022-01776-8}{Few-Body Syst. \textbf{63}, 74 (2022).}


\bibitem{Bilel20222} B. Hamil, B. C. L\"{u}tf\"{u}o\u{g}lu, \href{https://doi.org/10.1140/epjp/s13360-022-03055-1}{Eur. Phys. J. Plus \textbf{137}, 812 (2022).}

\bibitem{Bilel20223} B. Hamil, B. C. L\"{u}tf\"{u}o\u{g}lu, \href{https://doi.org/10.1140/epjp/s13360-022-03463-3}{Eur. Phys. J. Plus \textbf{137}, 1241 (2022).}

\bibitem{Halberg2022} A. Schulze-Halberg, \href{https://doi.org/10.1088/1402-4896/ac807a}{Phys. Scr. \textbf{97}, 085213 (2022).}

\bibitem{Samira2022} S. Hassanabadi, J.K\v{r}\'{\i}\v{z}, B. C. L\"{u}tf\"{u}o\u{g}lu, H. Hassanabadi, \href{https://doi.org/10.1088/1402-4896/aca2f7}{Phys. Scr. \textbf{97}, 125305 (2022).}


\bibitem{Najafizade1} A. Najafizade, H. Panahi, W. S. Chung, H. Hassanabadi, \href{https://doi.org/10.1063/5.0041830}{J. Math. Phys. \textbf{63}, 033505 (2022).}

\bibitem{Najafizade2} A. Najafizade, H. Panahi, \href{https://doi.org/10.1142/S0217732322500237}{Mod. Phys. Lett. A  \textbf{37}, 2250023 (2022).}

\bibitem{Mota20221} R. D. Mota, D. Ojeda-Guill\'{e}n,  \href{https://doi.org/10.1142/S0217732322502248}{Mod. Phys. Lett. A \textbf{37},  2250224 (2022).}


\bibitem{Ghazouani2023} S. Ghazouani, \href{https://doi.org/10.1088/1751-8121/acad4b}{J. Phys. A: Math. Theor. \textbf{55}, 505203 (2023).}

\bibitem{ChungEPL2023} W. S. Chung, M. de Montigny, H. Hassanabadi, \href{https://doi.org/10.1209/0295-5075/acc352}{ EPL \textbf{141}, 60004 (2023).}

\bibitem{Dong2023} S. H. Dong, L. F. Quezada, W. S. Chung, P. Sedaghatnia, H. Hassanabadi, \href{https://doi.org/10.1016/j.aop.2023.169259}{ Ann. Phys. \textbf{451}, 169259 (2023).}

\bibitem{Junker2023} G. Junker, S. H. Dong, P. Sedaghatnia, W. S. Chung, H. Hassanabadi, \href{https://doi.org/10.1016/j.aop.2023.169336}{
Ann. Phys. \textbf{454}, 169336 (2023).}

\bibitem{Sedaghatnia2023} P. Sedaghatnia, H. Hassanabadi, G. Junker, J. K\u{r}\'i\u{z}, W. S. Chung, \href{https://doi.org/10.1016/j.aop.2023.169445}{Ann. Phys. \textbf{458}, 169445 (2023).}


\bibitem{Rouabhia2023} N. Rouabhia, M. Merad, B. Hamil, \href{https://doi.org/10.1209/0295-5075/acf409}{EPL \textbf{143}, 52003 (2023).}



\bibitem{Samira2023} S. Hassanabadi, P. Sedaghatnia, W. S. Chung, B. C. L\"{u}tf\"{u}o\u{g}lu, J.K\v{r}\'{\i}\v{z},  H. Hassanabadi, \href{https://doi.org/10.1140/epjp/s13360-023-03933-2}{Eur. Phys. J. Plus \textbf{138}, 331 (2023).} 



\bibitem{Fateh20231} F. Merabtine, B. Hamil, B. C. L\"{u}tf\"{u}o\u{g}lu, A. Hocine, M. Benarous, \href{https://doi.org/10.1088/1742-5468/acd106}{J. Stat. Mech.  \textbf{2023}, 053102 (2023). }

\bibitem{Fateh20232} F. Merabtine, B. Hamil, B. C. L\"{u}tf\"{u}o\u{g}lu, \href{https://doi.org/10.1016/j.physa.2023.128841}{Physica A \textbf{623}, 128841 (2023). }





\bibitem{Quesne2023} C. Quesne, \href{https://doi.org/10.1088/1751-8121/acd736}{ J. Phys. A: Math. Theor. \textbf{56}, 265203 (2023).}

\bibitem{Quesne2024} C. Quesne, \href{https://doi.org/10.1209/0295-5075/ad2947}{EPL \textbf{145}, 62001 (2024).}

\bibitem{Mota20241} R. D. Mota, D. Ojeda-Guill\'{e}n,  M. A. Xicot\'encatl, \href{https://doi.org/10.1016/j.physa.2024.129525}{Physica A \textbf{635},  129525
(2024).}

\bibitem{Mota20242} R. D. Mota, D. Ojeda-Guill\'{e}n,  M. A. Xicot\'encatl, \href{https://doi.org/10.1007/s00601-024-01898-1}{Few-Body Syst. \textbf{65},  25
(2024).}

\bibitem{Schulze20242} A. Schulze-Halberg, \href{https://doi.org/10.1007/s00601-024-01931-3}{Few-Body Syst. \textbf{65}, 58 (2024).}

\bibitem{Junker2024} G. Junker, \href{https://doi.org/10.1088/1751-8121/ad213d}{J. Phys. A: Math. Theor. \textbf{57}, 075201 (2024).}

\bibitem{Schulze20241} A. Schulze-Halberg, P. Roy, \href{https://doi.org/10.1088/1751-8121/ad48eb}{J. Phys. A: Math. Theor. \textbf{57}, 225204 (2024).}



\bibitem{Schulze20243} A. Schulze-Halberg, \href{https://doi.org/10.1142/S0217751X24500131}{Int. J. Mod. Phys. A \textbf{39}, 2450013 (2024).}


\bibitem{Benzair2024} H. Benzair, T. Boudjedaa, M. Merad, \href{https://doi.org/10.1088/1402-4896/ad39b7}{Phys. Scr. \textbf{99}, 055261 (2024). }



\bibitem{Ballesteros2024} A. Ballesteros, A. Najafizade, H. Panahi, H. Hassanabadi, S. H. Dong, \href{https://doi.org/10.1016/j.aop.2023.169543}{Ann. Phys. \textbf{460}, 169543 (2024).}

\bibitem{Hocine20241} A. Hocine, B. Hamil, F. Merabtine,  B. C. L\"{u}tf\"{u}o\u{g}lu,  M. Benarous, \href{https://doi.org/10.31349/RevMexFis.70.051701}{Rev. Phys. Mex.  \textbf{70}, 051701 (2024).}



\bibitem{Bouguerne2024} H. Bouguerne, B. Hamil, B. C. L\"{u}tf\"{u}o\u{g}lu, M. Merad,   \href{ https://doi.org/10.1007/s12648-024-03170-y} {Ind. J. Phys. \textbf{98}, 4093 (2024).}

\bibitem{Hocine20242} A. Hocine, F. Merabtine, B. Hamil, B. C. L\"{u}tf\"{u}o\u{g}lu, M. Benarous,  \href{https://doi.org/10.1007/s12648-024-03311-3}{Ind. J. Phys. \textbf{in press},  (2024).}

\bibitem{Benchikha20241} A. Benchikha, B. Hamil, B. C.  L\"{u}tf\"{u}o\u{g}lu, B. Khantoul, \href{https://doi.org/10.1088/1402-4896/ad7aaf}{Phys. Scr.  \textbf{99}, 105274 (2024).}.

\bibitem{Benchikha20242} A. Benchikha, B. Khantoul, B. Hamil, B. C.  L\"{u}tf\"{u}o\u{g}lu, 
\href{https://doi.org/10.1007/s10773-024-05786-6}{Int. J. Theor. Phys. \textbf{63}, 248 (2024).}

\bibitem {Dunkl1989} C. F. Dunkl, \href{https://www.ams.org/journals/tran/1989-311-01/S0002-9947-1989-0951883-8/S0002-9947-1989-0951883-8.pdf}{Trans. Am. Math. Soc. \textbf{311}, 167 (1989).}

\bibitem {Wigner1950} E. P. Wigner, \href{https://doi.org/10.1103/PhysRev.77.711}{Phys. Rev. \textbf{77}, 711 (1950).}

\bibitem {Yang1951} L. M. Yang, \href{https://doi.org/10.1103/PhysRev.84.788}{Phys. Rev. \textbf{84}, 788 (1951).}

\bibitem{Green1953} H. S. Green, \href{https://doi.org/10.1103/PhysRev.90.270}{ Phys. Rev. \textbf{90}, 270 (1953).}

\bibitem{Greenberg1964} O. W. Greenberg, \href{https://doi.org/10.1103/PhysRevLett.13.598}{Phys. Rev. Lett. \textbf{13}, 598 (1964). }

\bibitem {Dunkl2014} C. F. Dunkl and Y. Xu, Encyclopedia of Mathematics and its Applications : Orthogonal Polynomials of Several Variables (Cambridge : Cambridge University Press) p 155 (2014).


\bibitem{Chakra1994} R. Chakrabarti, R. Jagannathan, \href{https://doi.org/10.1088/0305-4470/27/9/007}{J. Phys. A: Math Gen \textbf{27}, L227 (1994).}

\bibitem{Mikn1} M. S. Plyushchay, \href{https://doi.org/10.1016/0370-2693(94)90828-1}{Phys. Lett. B \textbf{320}, 91 (1994). }

\bibitem{Mikn2} M. S. Plyushchay, \href{https://doi.org/10.1016/0370-2693(94)90828-1}{Ann. Phys. \textbf{245}, 339 (1996). }

\bibitem {Hikami1996} K. Hikami, \href{https://doi.org/10.1143/JPSJ.65.394}{J. Phys. Soc. Japan 65, 394 (1996).}


\bibitem{Lapointe1996} L. Lapointe, L. Vinet, Commun. \href{https://doi.org/10.1007/BF02099456}{Math. Phys. \textbf{178}(2), 425 (1996).}

\bibitem{Kakei1996} S. Kakei, \href{https://doi.org/10.1088/0305-4470/29/24/002}{J. Phys. A \textbf{29}, L619 (1996).}

\bibitem{Mik1} M. Plyushchay, \href{https://doi.org/10.1016/S0550-3213(97)00065-5}{Nucl. Phys. B \textbf{491}, 619 (1997).}

\bibitem{Gamboa1999} J. Gamboa, M. Plyushchay, J. Zanelli, \href{https://doi.org/10.1016/S0550-3213(98)00832-3}{Nucl. Phys. B \textbf{543}, 447 (1999). }

\bibitem{Mik2} M. Plyushchay, \href{https://doi.org/10.1142/S0217751X0000198X}{Int. J. Mod. Phys. A \textbf{15}, 3679 (2000).}

\bibitem{Klishevich2001} S. M. Klishevich, M. Plyushchay, M. Rausch de Traubenberg, \href{https://doi.org/10.1016/S0550-3213(01)00442-4}{Nucl. Phys. B \textbf{616}, 419 (2001).}


\bibitem{Mikn3} P. A. Hortv\'athy, M. S. Plyushchay, \href{https://doi.org/10.1016/j.physletb.2004.05.043}{Phys. Lett. B \textbf{595}, 547 (2004). }

\bibitem{Rodrigues2009} R. de Lima Rodrigues, \href{https://doi.org/10.1088/1751-8113/42/35/355213}{J. Phys.  A \textbf{42}, 355213 (2009).}

\bibitem{Horvathy2010} P. A. Hortv\'athy, M. Plyushchay, M. Valenzuela, \href{https://doi.org/10.1016/j.aop.2010.02.007}{Ann. Phys. \textbf{325}, 1931 (2010).}



\bibitem{Mikn4} F. Correa, O. Lechtenfeld, M. S. Plyushchay, \href{https://doi.org/10.1007/JHEP04(2014)151}{J. High Energy Phys. \textbf{2014}, 151 (2014).}

\bibitem {Ubriaco2014} M. R. Ubriaco, \href{https://doi.org/10.1016/j.physa.2014.06.087}{Physica A \textbf{414}, 128 (2014).}

\bibitem {Jang2016} E. J. Jang, S. Park, W. S. Chung, \href{https://doi.org/10.3938/jkps.68.379}{J. Korean Phys. Soc. \textbf{68}, 379 (2016).}

\bibitem{Luo2020} Y. Luo, S. Tsujimoto, L. Vinet, A. Zhedanov, \href{https://doi.org/10.1088/1751-8121/ab63a9}{J. Phys. A: Math. Theor. \textbf{53}, 085205 (2020).}

\bibitem{Chung20202} W. S. Chung, H. Hassanabadi, \href{https://doi.org/10.31349/RevMexFis.66.308}{Rev. Mex. Fis. \textbf{66}, 308 (2020).}


\bibitem {Chung2021} W. S. Chung, H. Hassanabadi, \href{https://doi.org/10.1142/S0217732321501273}{Mod. Phys. Lett. A \textbf{36}, 2150127 (2021).} 

\bibitem{Ghazouani2021} S. Ghazouani, \href{https://doi.org/10.1007/s13324-020-00470-4}{Anal. Math. Phys. \textbf
{11}, 35 (2021).}

\bibitem{AHalberg2022} A. Schulze-Halberg, \href{https://doi.org/10.1142/S0217732322501784}{Mod. Phys. Lett. A \textbf{37}, 2250178 (2022).}

\bibitem {MotOjeda2022} R. D. Mota, D. Ojeda-Guill\'{e}n, \href{https://doi.org/10.1142/S0217732322500067}{
Mod. Phys. Lett. A \textbf{37}, 2250006 (2022).}


\bibitem {Dong2022} S. H. Dong, W. S. Chung, G. Junker, H. Hassanabadi, \href{https://doi.org/10.1016/j.rinp.2022.105664}{Results Phys. \textbf{39}, 105664 (2022).}

\bibitem {Sedaghatnia20232} P. Sedaghatnia, H. Hassanabadi, A. D. Alhaidari, W. S. Chung, \href{https://doi.org/10.1142/S0217751X22502232}{Int. J. Mod. Phys. A \textbf{37}, 2250223 (2023).}

{
\bibitem{nef1} W. S. Chung, G. Junker, S. H. Dong, H. Hassanabadi, \href{https://doi.org/10.1209/0295-5075/acaf9e}{EPL \textbf{141}, 32001 (2023).}

\bibitem{nef7} S. Hassanabadi,  J. K\u{r}\'i\u{z}, B. C. L\"{u}tf\"{u}o\u{g}lu, W. S. Chung, P.~Sedaghatnia, H. Hassanabadi, \href{https://doi.org/10.1007/s10773-024-05862-x} {Int. J. Theor. Phys.  \textbf{63}, 323 (2024).}

\bibitem{nef8} D. Nath, N. Ghosh, A. K. Roy, \href{https://doi.org/10.1063/5.0200405}{J. Math. Phys. \textbf{65}, 083511 (2024).}
}


\bibitem{Hartman1972} H. Hartmann, \href{https://doi.org/10.1007/BF00641399}{Theoret. Chim. Acta \textbf{24}, 201 (1972).} 


\bibitem{Hautot1973} A. Hautot, \href{https://doi.org/10.1063/1.1666184}{J. Math. Phys. \textbf{14}, 1320 (1973).}

\bibitem{Kibler1984}  M. Kibler, T. Négadi, \href{ https://doi.org/10.1002/qua.560260308}{Int. J. Quantum Chem. \textbf{26}, 405 (1984).}

 \bibitem{Kibler1987}  M. Kibler, T. Négadi, \href{ https://doi.org/10.1002/qua.560260308}{Phys. Lett. A \textbf{124}, 42 (1987).}

 \bibitem{Quesne1988} C. Quesne, \href{https://doi.org/10.1088/0305-4470/21/14/010}{J. Phys. A: Math. Gen. \textbf{21}, 3093 (1988).}  

\bibitem{Draganascu1992} Gh. E. Draganascu, C. Campigotto, M. Kibler, \href{https://doi.org/10.1016/0375-9601(92)90883-N}{Phys. Lett. A \textbf{170}, 339 (1992).}

\bibitem{Chang1999} M. H. Chang, T. Chiueh, C. R. Lo, \href{https://doi.org/10.1103/PhysRevE.59.67}{Phys. Rev. E \textbf{59}, 67 (1999).}


\bibitem{Chen2002} C. Y. Chen, C. L. Liu, D. S. Sun, \href{https://doi.org/10.1016/s0375-9601(02)01477-9}{Phys. Lett. A, \textbf{305}, 341 (2002).}


\bibitem{Alhaidari2005} A. D. Alhaidari, \href{https://doi.org/10.1088/0305-4470/38/15/012}{J. Phys. A: Math. Gen. \textbf{38}, 3409 (2005).}

{
\bibitem{nef2} C. Y. Chen, S. H. Dong, \href{https://doi.org/10.1016/j.physleta.2004.12.062}{Phys. Lett. A \textbf{335}, 374 (2005).}

\bibitem{nef3} S. H. Dong, G. H. Sun, M. Lozada-Cassou, \href{https://doi.org/10.1016/j.physleta.2005.04.024}{Phys. Lett. A \textbf{340}, 94 (2005).}
}


\bibitem {Berkdemir2008} C. Berkdemir,  R. Sever,  \href{https://doi.org/10.1088/1751-8113%2F41%2F4%2F045302}{J. Phys. A: Math. Theor. \textbf{41}, 045302 (2008).}

\bibitem {Berkdemir20091} C. Berkdemir, R. Sever, \href{https://doi.org/10.1007/s10910-008-9447-7}{J. Math. Chem. \textbf{46}, 139 (2009).}

\bibitem {Berkdemir20092} C. Berkdemir, R. Sever, \href{https://doi.org/10.1007/s10910-008-9498-9}{J. Math. Chem. \textbf{46}, 1122 (2009).}

\bibitem{Zhang2010} M. C. Zhang, G. H. Sun, S. H. Dong, \href{https://doi.org/10.1016/j.physleta.2009.11.072}{Phys. Lett. A \textbf{374}, 704 (2010).}

\bibitem{Agboola2011} D. Agboola, \href{https://doi.org/10.1088/0253-6102/55/6/06}{Commun. Theor. Phys. \textbf{55}, 972 (2011).} 

{
\bibitem{nef4} C. Y. Chen, Y. You, X. H. Wang, S. H. Dong, \href{https://doi.org/10.1016/j.physleta.2013.04.026}{Phys. Lett. A \textbf{377}, 1521 (2013).}

\bibitem{nef5} C. Y. Chen, F. A. Lu, D. S. Sun, S. H. Dong, \href{https://doi.org/10.1088/1674-1056/22/10/100302}{Chinese Phys. B \textbf{22}, 100302 (2013).}

\bibitem{nef6} D. S. Sun, F. L. Lu, C. Y. Chen, S. H. Dong, \href{https://doi.org/10.1142/S0217732315502004}{Mod. Phys. Lett. A \textbf{30}, 1550200 (2015).}
}


\bibitem{Afs2015} A. Afshardoost, H. Hassanabadi, \href{https://doi.org/10.1139/cjp-2015-0520}{Can. J. Phys. \textbf{94}, 71 (2016).}


\bibitem{Li2017} W. Li, C. Y. Chen, S. H. Dong, \href{https://doi.org/10.1155/2017/7374256}{Adv. High Energy Phys. \textbf{2017}  7374256, (2017).}


\bibitem{You2018} Y. You, F. L. Lu, D. S. Sun, C. Y. Chen, S. H. Dong, \href{https://doi.org/10.1155/2018/5824271}{Adv. High Energy Phys. \textbf{2018}  8307486, (2018).}

\bibitem{Sobhani2019} H. Sobhani, H. Hassanabadi, \href{https://doi.org/10.1016/j.nuclphysa.2019.121621}{Nucl. Phys. A \textbf{992}, 121621 (2019).} 

\bibitem{Moumni2020} M. Moumni, M. Falek, M. Heddar, \href{https://doi.org/10.1007/s00601-020-01580-2}{Few-Body Syst. \textbf{61}, 47 (2020).}  

\bibitem{Bouchefra2022} D. Bouchefra, B. Boudjedaa, \href{https://doi.org/10.1140/epjp/s13360-022-02976-1}{Eur. Phys. J. Plus \textbf{137}, 743 (2022).}  

\bibitem{Ahmadov2023} A. I. Ahmadov, M. Demirci, M. F. Mustamin, M. Sh. Orujova,   \href{https://doi.org/10.1140/epjp/s13360-023-03715-w}{Eur. Phys. J. Plus \textbf{138}, 92 (2023).}  

\bibitem{Arabsaghari2024} A. Arabsaghari, H. Hassanabadi, W. S. Chung, \href{Accepted}{Accepted Mod. Phys. Lett. A \textbf{xx}, xx (2024).}

\bibitem{Lewis1969} H. R. Lewis, W. B. Riesenfeld, \href{https://doi.org/10.1063/1.1664991}{J. Math. Phys. \textbf{1010}, 1458 (1969).}







\bibitem {Ang3} A. A. Makarov, J. A. Smorodinsky, K. Valiev, P. Winternitz. \href{https://doi.org/10.1007/BF02755212}{Nuovo Cime. A  \textbf{52}, 1061 (1967). } 



\bibitem {Ang4} S. H. Dong, C. Y. Chen, M.  Lozada-Cassou,  \href{https://doi.org/10.1002/qua.20729}{Int. J. Quantum Chem. \textbf{105}, 453 (2005).
} 

\bibitem {Ang5}  A. F. Nikiforov, V. B. Uvarov,  \emph{Special functions of mathematical physics \textbf{205}, } (Birkhuser, Basel, 1988).

\bibitem {Ang6} C. Tezcan, R. Sever, \href{https://doi.org/10.1007/s10773-008-9806-y}{Int. J. Theor. Phys. \textbf{48}, 337 (2009).}

\bibitem {Ang7} R. Koekoek,  P. A. Lesky,  R. F. Swarttouw,  \emph{Hypergeometric Orthogonal Polynomials and Their q-Analogues}, (Springer,  Heidelberg, 2010).





\end{thebibliography}
\end{document}